\documentclass[ aps, prl, floatfix, preprintnumbers, reprint,
superscriptaddress, amsmath, amssymb, longbibliography, ]{revtex4-2}%
\usepackage{float}
\usepackage{soul}
\usepackage[utf8]{inputenc}
\usepackage{graphicx}
\usepackage{xcolor}
\usepackage{bm}
\usepackage{xr}
\usepackage[breaklinks=true]{hyperref}
\usepackage{mathtools}
\usepackage{amsfonts}
\usepackage{amssymb}
\usepackage{amsmath}%
\providecommand{\U}[1]{\protect\rule{.1in}{.1in}}

\hypersetup{bookmarksnumbered, pdfpagemode=UseOutlines,
	pdfauthor={F.\ Feringa},
	pdftitle={Observation of magnetization surface textures of the van der Waals antiferromagnet FePS$_{3}$ by spin Hall magnetoresistance},
	pdfdisplaydoctitle,
	colorlinks=true, citecolor=blue, filecolor=blue, linkcolor=blue, urlcolor=blue}
\begin{document}
\title{Observation of magnetization surface textures of the van der Waals antiferromagnet FePS$_{3}$ by spin Hall magnetoresistance}

\author{F.\ \surname{Feringa}}
\email[e-mail: ]{F.Feringa@rug.nl}
\affiliation{Physics of Nanodevices, Zernike Institute for Advanced Materials, University of Groningen, 9747 AG Groningen, The Netherlands}
\author{G.\ E.\ W.\ \surname{Bauer}}
\affiliation{Physics of Nanodevices, Zernike Institute for Advanced Materials, University of Groningen, 9747 AG Groningen, The Netherlands}
\affiliation{AIMR \& Institute for Materials Research, Tohoku University, Aoba-ku, Katahira 2-1-1, Sendai, Japan}
\author{B.\ J.\ \surname{van Wees}}
\email[e-mail: ]{B.J.van.Wees@rug.nl}
\affiliation{Physics of Nanodevices, Zernike Institute for Advanced Materials, University of Groningen, 9747 AG Groningen, The Netherlands}

\begin{abstract}
Van der Waals materials are a new platform to study two-dimensional
systems, including magnetic order. Since the number of spins is relatively
small, measuring the magnetization is challenging. Here we  report spin Hall magnetoresistance (SMR) up to room temperature caused by the magnetic surface texture of exfoliated flakes of magnetic van der Waals materials. For the antiferromagnet FePS$_{3}$ the SMR amounts to 0.1 $\%$
for an applied magnetic field of 7 T at 5 K which implies a substantial canting of the magnetic moments relative to the colinear antiferromagnetic order. The canting is substantial even for a magnetic field along the N\'{e}el vector, which illustrates the unique power of the SMR to detect magnetic surface textures in van der Waals magnets. \\ \\

\end{abstract}
\maketitle


\affiliation{Physics of Nanodevices, Zernike Institute for Advanced Materials, University of Groningen, 9747 AG Groningen, The Netherlands}

\affiliation{Physics of Nanodevices, Zernike Institute for Advanced Materials, University of Groningen, 9747 AG Groningen, The Netherlands}
\affiliation{AIMR \& Institute for Materials Research, Tohoku University, Aoba-ku, Katahira 2-1-1, Sendai, Japan}

\affiliation{Physics of Nanodevices, Zernike Institute for Advanced Materials, University of Groningen, 9747 AG Groningen, The Netherlands}


\affiliation{Physics of Nanodevices, Zernike Institute for Advanced Materials, University of Groningen, 9747 AG Groningen, The Netherlands}

\affiliation{Physics of Nanodevices, Zernike Institute for Advanced Materials, University of Groningen, 9747 AG Groningen, The Netherlands}
\affiliation{AIMR \& Institute for Materials Research, Tohoku University, Aoba-ku, Katahira 2-1-1, Sendai, Japan}

\affiliation{Physics of Nanodevices, Zernike Institute for Advanced Materials, University of Groningen, 9747 AG Groningen, The Netherlands}


\affiliation{Physics of Nanodevices, Zernike Institute for Advanced Materials, University of Groningen, 9747 AG Groningen, The Netherlands}

\affiliation{Physics of Nanodevices, Zernike Institute for Advanced Materials, University of Groningen, 9747 AG Groningen, The Netherlands}
\affiliation{AIMR \& Institute for Materials Research, Tohoku University, Aoba-ku, Katahira 2-1-1, Sendai, Japan}

\affiliation{Physics of Nanodevices, Zernike Institute for Advanced Materials, University of Groningen, 9747 AG Groningen, The Netherlands}


The magnetic susceptibility is a crucial material property that can be
measured by many techniques, such as vibrating sample, torque, or squid
magnetometry \cite{Franco2021}. Since signals are proportional to the total
magnetization, these techniques are not sensitive to the relatively small number
of spins at the surface of bulk crystals or in two-dimensional van der Waals materials.

The spin Hall magnetoresistance (SMR) in heavy metal contacts to (preferably
electrically insulating) ferromagnets senses only the spin configuration at the interface
and is therefore well suited to study the magnetic properties of the surfaces
of bulk magnets or van der Waals (vdW) magnets \cite{Burch2018,Gibertini2019} down to monolayer
thicknesses. The SMR is governed by the exchange interaction $\sim\vec{\sigma
}\cdot\vec{m}$ of the spins of conduction electrons with polarization
$\vec{\sigma}$ in the heavy metal and the local moment direction $\vec{m}$ in
the magnet \cite{Nakayama2013,Chen2013,Vlietstra2013,Vlietstra2013a}. The
resistance of Pt contacts to ferromagnetic insulators is lowest when
$\vec{\sigma}\Vert\vec{m}$ by combination of the direct and inverse spin Hall
effects. In antiferromagnets the SMR may become \textquotedblleft
negative\textquotedblright, i.e. a\ minimum resistance when the N\'{e}el vector is parallel to $\vec{\sigma}$ $(\vec{\sigma}\Vert\vec{n})$; note that in the absence of in-plane anisotropy the N\'{e}el vector orients normal to the applied magnetic field due to the minimization of the exchange and Zeeman energy \cite{Hoogeboom2017,Fischer2018}. SMR has been observed in paramagnetic
CoCr$_{2}$O$_{4}$ \cite{Aqeel2015a}, amorphous YIG \cite{Lammel2019}, Cr$_{2}%
$O$_{3}$ \cite{Schlitz2018}, NdGaO$_{3}$ \cite{Eswara}, and gadolinium gallium garnet (GGG)
\cite{Oyanagi2020}. The SMR revealed magnetic phase transitions in $\alpha
$-Fe$_{2}$O$_{3}$ \cite{Lebrun2019}, CoCr$_{2}$O$_{4}$ \cite{Aqeel2015a}, and
DyFeO$_{3}$ \cite{Hoogeboom2021} and magnetic domain structures
\cite{Damerio2021}.

\begin{figure}[bh]
\centering
\includegraphics[width=1\linewidth]{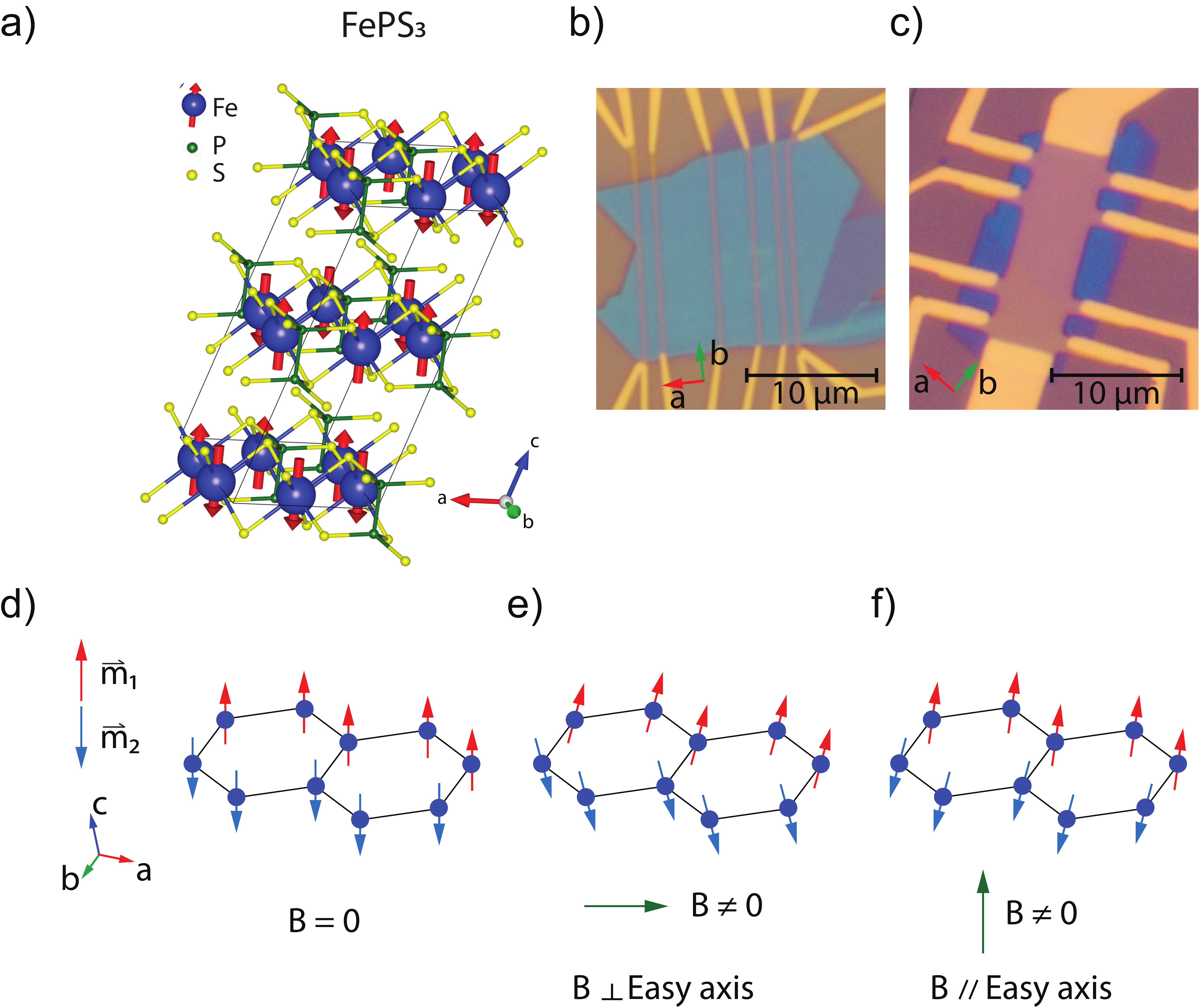} \caption{a) Crystal
structure of FePS$_{3}$ generated by VESTA \cite{Momma2011}. The red arrows
are the directions of the Fe$^{2+}$ magnetic moments at zero magnetic field
with the easy axis normal to the $\mathbf{a}$-$\mathbf{b}$ crystal plane. b) and
c) are the experimental samples imaged by optical microcopy. d-f) Schematic
response of the magnetic sublattices $m^{(1)}$ and $m^{(2)}$ without and with
magnetic field parallel and perpendicular to the $\mathbf{a}$-$\mathbf{b}$
plane, respectively. A magnetic field along $\mathbf{a}$, normal to the easy
axis, generates a transverse magnetization, while a magnetic field along the easy axis
causes an unconventional canting also in the $\mathbf{a}$ direction.}%
\label{fig:image1}%
\end{figure}

\begin{figure*}[th]
\hspace*{-1.8cm} \centering
\includegraphics[width=210mm]{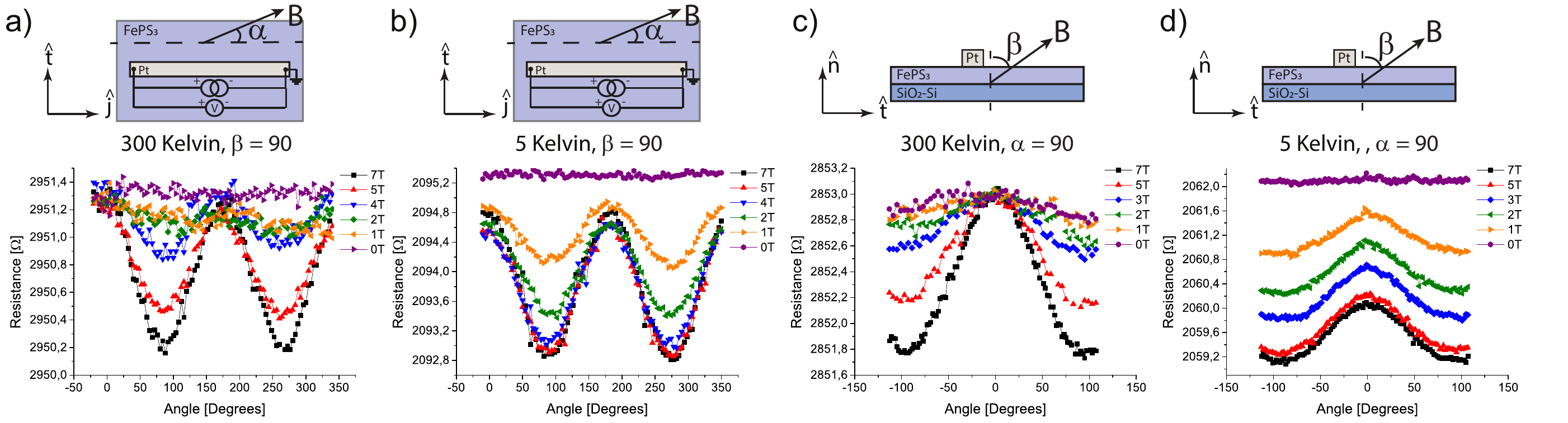} \caption{SMR
response for an in-plane and out-of-plane rotation of an external magnetic
field at 300 and 5 Kelvin at various magnetic field strengths. The decrease of
the background resistance at 5 kelvin between 0 and 1 tesla in b) and d) is attributed to
suppression of weak localization effects in Pt. }%
\label{fig:image2}%
\end{figure*}

The SMR in the longitudinal resistance of a metal contact on an
antiferromagnet with two sublattices $A$ and $B$ reads
\begin{equation}
\mathrm{SMR}=\rho_{L}-\rho_{0}\,=\frac{1}{2}\sum_{X=A,B}\rho_{1}^{\left(
X\right)  }\left[  1\,-(\vec{m}^{\left(  X\right)  }\,\cdot\,\vec{\sigma}%
)^{2}\right]  \label{rhoL}%
\end{equation}
where $\rho_{1}^{\left(  X\right)  }$ is the SMR coefficient for the
sublattice $X$ with local moment magnetization $\vec{m}^{\left(  X\right)  }$
controlled by an applied magnetic field,\thinspace while $\vec{\sigma}$ is the
current-induced spin accumulation direction generated by the spin Hall effect
in the current-biased contact \cite{Fischer2018,Geprags2020}. 

Here we report for the first time
observation of an SMR on the exfoliated van der Waals material FePS$_{3}$, an
uniaxial antiferromagnet with perpendicular magnetic anisotropy. The SMR persists above the N\'{e}el temperature in the paramagnetic regime up to 300 kelvin. We discover
an unusual symmetry breaking of the surface magnetic textures by out-of-plane
magnetic fields that cannot be accessed by conventional magnetometry. 

\begin{figure}[ptb]
\hspace*{-5mm} \centering
\includegraphics[width = 100mm]{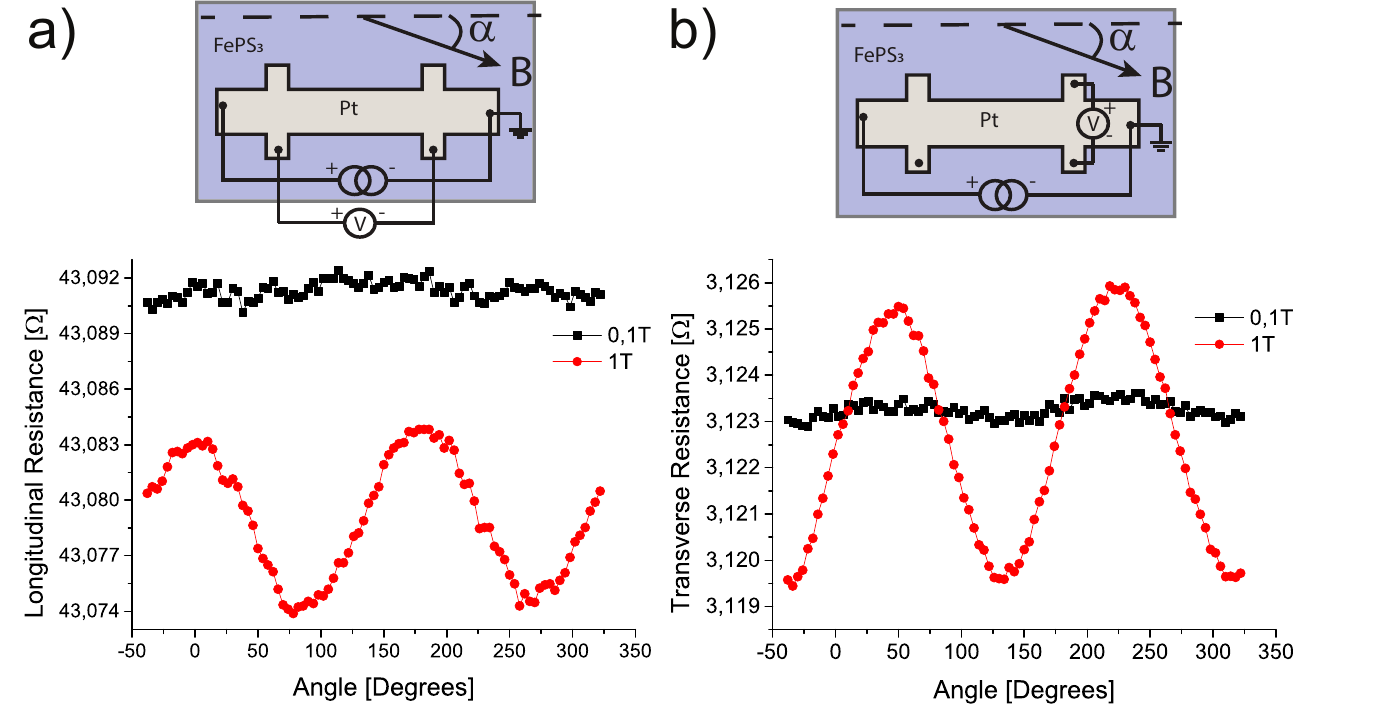} \caption{The
a) longitudinal and b) transverse response to a rotating magnetic field of 0.1
and 1 Tesla for a current density of 11\thinspace$\times$ 10$^{-14}$\thinspace
A/m$^{2}$ at 10 kelvin. The sign corresponds to a positive SMR and the ratio of the amplitudes agrees with the aspect ratio of 1.6 of the Hall bar. The decrease of the background resistance between 0.1 and 1 tesla is
attributed to suppression of weak localization effects in Pt, similar to what
is observed in Fig. \ref{fig:image2}. }%
\label{fig:image2_2}%
\end{figure}

\begin{figure*}[t]
\hspace*{-1.8cm} \centering
\includegraphics[width=210mm]{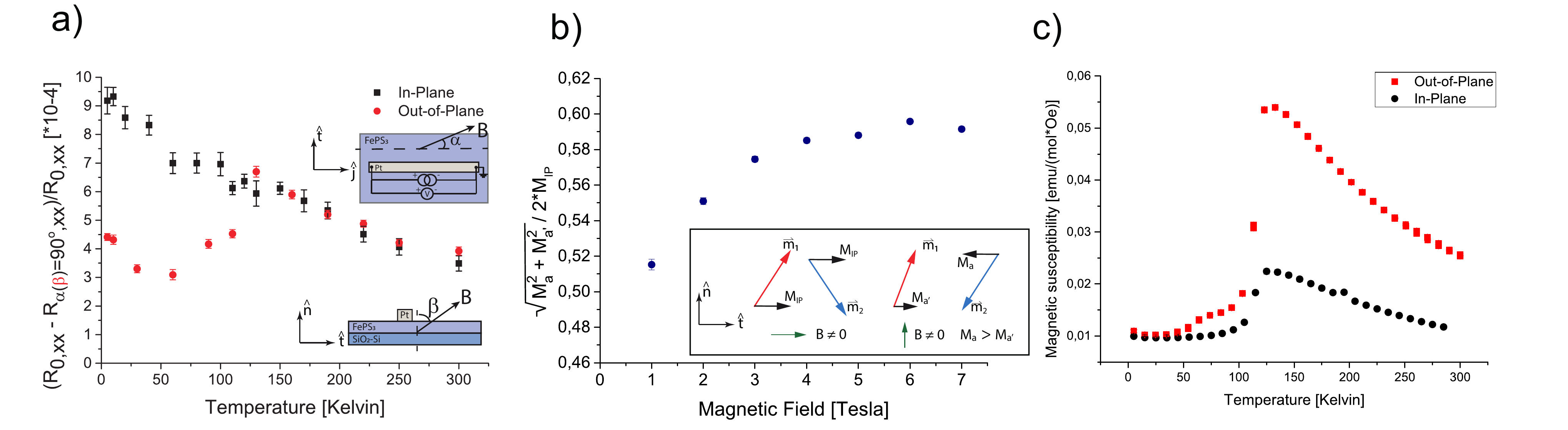} \caption{a) SMR signals
as a function of temperature for an in-plane and out-of-plane rotation of the
magnetic field with $\alpha=\beta=90{{}^{\circ}}$ and $B=7\,\mathrm{T.}$ b)
Ratio $\sqrt{M_{a}^{2}+M_{a^{\prime}}^{2}} / 2M_{IP}$ for various magnetic
field strengths at $T=5\,$K extracted by fitting Eq. (\ref{BetaRho}) to Fig.
\ref{fig:image2}d). Inset shows the definitions of $M_{IP}$, $M_{a^{\prime}}$
and $M_{a}$. c) The magnetic susceptibilities of bulk FePS$_{3}$ for in-plane
and out-of-plane magnetic fields. See \cite{Sup6} for more details.}%
\label{fig:image4}%
\end{figure*}

FePS$_{3}$ is a transition metal phosphorus trichalcogenide MPX$_{3}$, where M
is a transition metal and X is a chalcogenide, which all crystallize with
monoclinic space group C$\frac{2}{m}$ symmetry \cite{Ouvrard1985}. MnPS$_{3}$
and FePS$_{3}$ are uniaxial antiferromagnets, while NiPS$_{3}$ is an
easy-plane antiferromagnet. Below the N\'{e}el temperature $T_{N}%
=120\,\mathrm{K}$ \cite{Lee2016} two nearest neighbor moments in a FePS$_{3}$
layer order ferromagnetically and one nearest neighbor antiferromagnetically.
Neighboring layers in turn order antiferromagnetically \cite{Joy1992a}, as
shown in Fig. \ref{fig:image1}a). The bandgap of FePS$_{3}$ is 1.5 eV \cite{Wang2018a} and we find that FePS$_{3}$ effectively behaves as a good electric insulator \cite{Sup3}.

We mechanically exfoliated FePS$_{3}$ on a SiO$_{2}$-Si substrate and in a
nitrogen atmosphere by adhesive tapes \cite{Novoselov2005} from a bulk crystal
\cite{HQ}. We patterned Hall bars and strips on top of 20\,nm and 40\,nm thick FePS$_{3}$ flakes
by conventional electron beam lithography, followed by dc sputtering
deposition of $\approx$ 7\thinspace nm of Pt. The Pt strips (Hall bar) are 400
nm $\left(  5\,\mathrm{\mu m}\right)  $ wide and approximately
15$\,\mathrm{\mu m}$ $\left(  15\,\mathrm{\mu m}\right)  $ long. Ti-Au
(5-55nm) leads deposited by e-beam evaporation connect the Pt Hall bar and
strips to a chip carrier by AlSi wires. Fig. \ref{fig:image1}b) and c) shows
images of the experimental samples.

We applied AC currents $I_{rms}=100\,\mathrm{\mu A}-4\,\mathrm{mA}$ at
$\omega=8\,\mathrm{Hz}$ to the Pt strips and Hall bar and measured the
voltages over the longitudinal and transverse directions. The current
generates a spin accumulation at the interface that interacts with the surface
magnetization. The associated Joule heating generates a temperature gradient
in the magnet via the bulk spin Seebeck effect (SSE) that scales quadratically 
with the current. Lock-in amplifiers separately measure the first $R_{1\omega
}=V_{1\omega}/I$ and second $R_{2\omega}=V_{2\omega}/I^{2}$ harmonic response.
We also monitor the conduction of FePS$_{3}$ between two Pt
strips, thereby excluding parasitic current flow through FePS$_{3}$, as
shown in the supplementary material (SM) Fig. S3.

Here we focus on $R_{1\omega},$ i.e. the SMR as a function of external magnetic
field at in-plane and out-of-plane angles relative to the Pt contact as
summarized in Fig. \ref{fig:image2} for the strip and Fig. \ref{fig:image2_2}
for the Hall bar device. The field generates a net magnetization in the
antiferromagnet by canting the local moments, as shown in Fig.
\ref{fig:image1}e). Since the in-plane magnetic anisotropy is small, the
in-plane magnetization closely follows the magnetic field direction. When
rotating a field with constant strength from the hard in-plane to an easy
normal direction, the in-plane magnetization must decreases. Up to an applied
magnetic field of $\left\vert \vec{B}\right\vert =7$\thinspace T, we do not
observe a magnetic (such as a spin-flop) phase transition or saturation of the in-plane magnetization, consistent with \cite{Wildes2020}.
Fig. \ref{fig:image2_2}a) and b) display the transverse and longitudinal
response measured in the Hall bar. The observed \textquotedblleft
positive\textquotedblright\ SMR in Pt on FePS$_{3}$ is similar to that of
ferromagnets and a consequence of the strong out-of-plane anisotropy. Due to the out-of-plane sub lattice magnetization, the spin absorption is maximum for $\vec{B}\,=0$, resulting in $\,\vec{m}^{\left(  X\right)  }\cdot\,\vec{\sigma}=0$ and therefore a higher Pt resistance. An in-plane magnetic field generates an in-plane magnetization parallel to the magnetic field $\vec{m}^{\left(  X\right)  }\Vert\vec{B}\,$ as shown in Fig. \ref{fig:image1}e). As as consequence, the spin absorption is reduced, which lowers the Pt resistance when the applied magnetic field is parallel to the spin accumulation ($\vec{B}\Vert\vec{\sigma}$), resulting in a \textquotedblleft positive\textquotedblright\ SMR for fields up to 7\,T. 


Figure \ref{fig:image4}a) shows the SMR at various temperatures for an
in-plane and out-of-plane magnetic field of $\left\vert \vec{B}\right\vert
=7$\thinspace T. The SMR measured in the out-of-plane rotation of the magnetic field
displays a pronounced cusp at $T_{N}$, very similar to the separately measured
out-of-plane magnetic susceptibility in Fig. \ref{fig:image4}c), which is
typical for an antiferromagnetic phase transition. On the other hand, the
in-plane SMR increases monotonically with decreasing temperature. The in-plane
magnetic susceptibility in Fig. \ref{fig:image4}c) has a step at $T_{N},$ but
this feature is much weaker than the step in the out-of-plane susceptibility.

\begin{figure*}[th]
	\hspace*{-1.8cm} \centering
	\includegraphics[width=210mm]{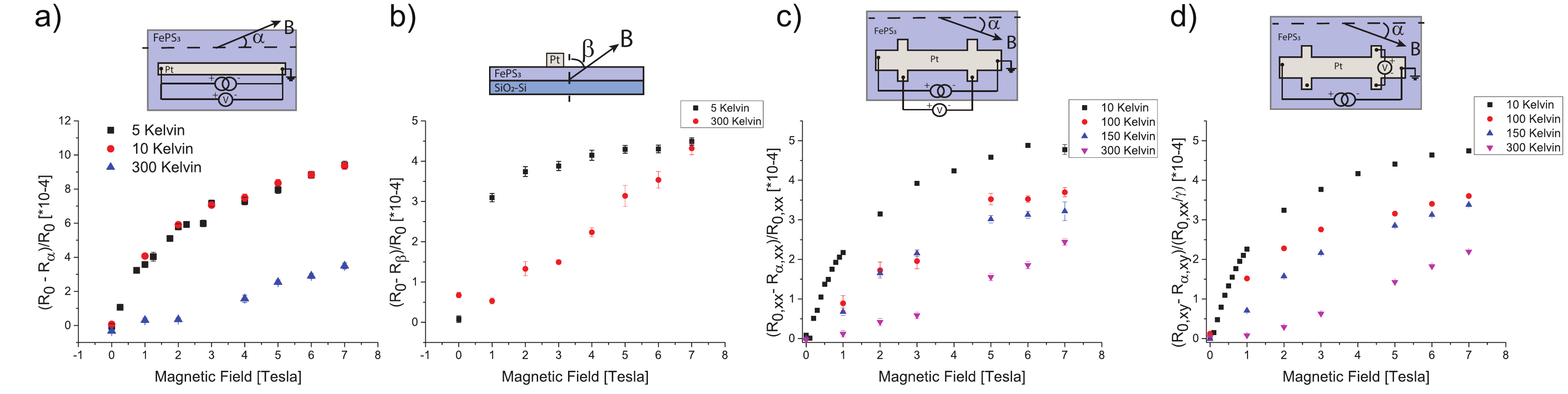} \caption{Magnetic
		field dependent SMR amplitudes for an a) in-plane and b) out-of-plane rotation
		of the magnetic field at temperatures indicated in the inset. The magnetic
		field dependent longitudinal and transverse SMR amplitudes for various
		temperatures are shown in c) and d). $\gamma$ = 1.6 is the geometric
		conversion factor which is the ratio between the length and the width of the
		Hall bar which is taken into account in order to compare the change in
		relative longitudinal resistance to the transverse results.}%
	\label{fig:image3}%
\end{figure*}

A spin model, which only includes uniaxial anisotropy, exchange, and Zeeman interaction, cannot describe the difference between the observed
in-plane and out-of-plane rotations below $T_{N}$ because the spin absorption should be maximum when the magnetic field is perpendicular to the spin accumulation (for $\alpha = 0^{\circ}$ and $\beta = 0^{\circ}$), i.e. $\vec{m}^{\left(  X\right)  }\,\cdot\,\vec{\sigma} = 0$ for $\vec{B}\,\cdot\,\vec{\sigma} = 0$.  Moreover, Fig. \ref{fig:image2}d) reveals an unexpected decrease of
the entire angle-dependent out-of-plane SMR at low temperatures.

The observations indicate that both sublattices 1 and 2$,$ originally parallel
and antiparallel to an out-of-plane magnetic field, cant into the
crystallographic $\mathbf{a}$ direction as sketched in Fig. \ref{fig:image1}f) and in the inset of Fig.
\ref{fig:image4}b). Since the canting occurs by a gain in the Zeeman energy,
the canting angle of sublattice 2 must be larger than that of $X=1$. For
$\alpha=90%
{{}^\circ}%
$ we may interpolate the projections of the sublattice magnetizations
$m^{\left(  X\right)  }(B,\beta)=\vec{m}^{\left(  X\right)  }\cdot\vec{\sigma
}$ as
\begin{subequations}
\label{mag1_2}%
\begin{align}
m^{\left(  1\right)  }(B,\beta)\, &  =M_{IP}(B)\sin\beta\ +M_{a%
}(B)\cos\beta\ \\
m^{\left(  2\right)  }(B,\beta)\, &  =M_{IP}(B)\sin\beta\ -M_{a^{\prime}}(B)\cos\beta
\end{align}
where $M_{IP}\ (M_{a^{\prime}}\ $and $M_{a}$) are the limiting values for
$\beta=90%
{{}^\circ}%
\left(  0%
{{}^\circ}%
\right)  $ and defined to be positive and $M_{a}>M_{a^{\prime}}$. This leads
to an SMR Eq. (\ref{rhoL}) %

\end{subequations}
\begin{equation} \label{BetaRho}
	\begin{aligned}
		\mathrm{SMR}(B,\beta)&/\rho_{1}=1-M_{IP}^{2}(B)\sin^{2}\beta \\
	&-\frac{1}{2}(M_{a}^{2}(B) + M_{a^{\prime}}^{2}(B))\cos^{2}\beta\\
	&-M_{IP}(M_{a}(B)-M_{a^{\prime}}(B))\sin\beta\cos\beta 
\end{aligned}
\end{equation}
and $\mathrm{SMR}(B,0%
{{}^\circ}%
)=\rho_{1}\left[  1-M_{IP}^{2}(B)\right]  $ and $\mathrm{SMR}(B,90%
{{}^\circ}%
)=\rho_{1}\left[  1-\frac{1}{2}(M_{a}^{2}(B)+M_{a^{\prime}}^{2}(B))\right]  $.

We estimate the ratio of the canting by an out-of-plane magnetic field relative to that of an in-plane
magnetic field, $\sqrt{M_{a}^{2}+M_{a^{\prime}}^{2}} / 2M_{IP}$, by fitting the observed SMR to Eq. (\ref{BetaRho}) for $\rho_{0}=2060\,\Omega$ and $\rho_{1}=2\,\Omega$ as a function of magnetic field
strength, see Fig. \ref{fig:image4}b. We arrive at ratios up to 0.6, which is
much larger than expected considering the high exchange and anisotropy energy
cost compared to the Zeeman energy gain. We note that the possibility of canting with a magnetic field parallel to the  N\'{e}el vector has been discussed in the context of observations of off-diagonal elements in the magnetic susceptibility matrix \cite{Nauman2021b}. 

Fig. \ref{fig:image3} displays the longitudinal and transverse SMR as a
function of magnetic field strengths and directions for selected temperatures. The magnetization of bulk FePS$_{3}$ increases linearly with magnetic field strength \cite{Sup6}, which according to Eq.\,(\ref{rhoL}), should result in a quadratic increase of the SMR amplitude. The observed linear increase with field is therefore an indication for different bulk and surface magnetizations.

Above the N\'{e}el temperature, SMR still exists but the difference between
the $\alpha$ and $\beta$ angular scans disappears in Fig. \ref{fig:image4},
therefore the effect of canting into the crystallographic $\mathbf{a}$ direction for an out-of-plane magnetic field disappears above $T_{N}$. The SMR
amplitude decreases with increasing temperature by thermal fluctuations.
Moreover, the second harmonic signal at room temperature\ is vanishingly small
in spite of the finite SMR. This points to a very short magnon diffusion
length or absence of long range order, since the bulk contribution dominates
the spin Seebeck effect \cite{Sup2}.

During finalization of this manuscript, Wu \textit{et al.} posted
an SMR study on the vdW antiferromagnet CrPS$_{4}$
\cite{wu}. This material has a very different spin
configuration and small magnetic anisotropy. Its spin-flop transition at weak
magnetic fields prohibits the canting we report here.

In conclusion, we succeeded in observing a strong SMR in exfoliated thin films
of the perpendicular antiferromagnetic van der Waals material FePS$_{3}$ with
Pt contacts. Below the N\'{e}el temperature, the SMR is very anisotropic at
least up to field strengths of 7\thinspace T. The difference between the SMR
and the field-induced magnetization of the bulk crystals is caused by a
difference in the magnetization of the uppermost layer that should be
indicative of the SMR in the monolayer limit. We observe a surprisingly large
collective canting\ of the spins in the crystallographic \textbf{a}
direction by an out-of-plane magnetic field. In the paramagnetic regime the
SMR signal of the field-induced magnetization loses the easy-plane anisotropy
observed in the antiferromagnetic phase. 

Our results reveal an unexpectedly complex field dependence of the
magnetization of a representative van der Waals antiferromagnet. The SMR does
not mirror the bulk magnetization because it senses only the uppermost
monolayer and we do not expect that results change much in the monolayer
limit. The technique can be used in principle for all van der Waals magnets
onto which Pt films can be deposited. \\ \\

	We acknowledge the technical support from J.\ G.\ Holstein, H.\ Adema T.\ Schouten and H. de Vries.
We acknowledge the financial support of the Zernike Institute for Advanced Materials.
This project has been supported by the NWO Spinoza prize awarded to Prof.\ B.\ J.\ van Wees by the NWO.	GB acknowledges fudning by JSPS Kakenhi grant \# 19H00645.	
\bibliographystyle{apsrev4-2}

\end{document}